%% file: icassp20d_submission.tex
\documentclass{article}
\usepackage{spconf,amsmath,graphicx}
\usepackage[moderate]{savetrees}
\usepackage{amssymb}
\usepackage{pifont}
\usepackage{multirow}
\usepackage{makecell}
\usepackage{listings}
\usepackage{xcolor}
\usepackage{dirtytalk}
\usepackage{booktabs}
\usepackage{hyperref}
\hypersetup{
    colorlinks=true,
    linkcolor=purple,
    filecolor=magenta,      
    urlcolor=purple,
}

\newcommand*\samethanks[1][\value{footnote}]{\footnotemark[#1]}
\usepackage[utf8]{inputenc}

\definecolor{codegreen}{rgb}{0,0.6,0}
\definecolor{codegray}{rgb}{0.5,0.5,0.5}
\definecolor{codepurple}{rgb}{0.58,0,0.82}
\definecolor{backcolour}{rgb}{0.95,0.95,0.92}
 
\lstdefinestyle{mystyle}{
    backgroundcolor=\color{backcolour},   
    commentstyle=\color{codegreen},
    keywordstyle=\color{magenta},
    numberstyle=\tiny\color{codegray},
    stringstyle=\color{codepurple},
    basicstyle=\ttfamily\footnotesize,
    breakatwhitespace=false,         
    breaklines=true,                 
    captionpos=b,                    
    keepspaces=true,                 
    numbers=left,                    
    numbersep=5pt,                  
    showspaces=false,                
    showstringspaces=false,
    showtabs=false,                  
    tabsize=2,
    basicstyle=\fontsize{7}{8}\ttfamily
}
 
\title{Disentangled Speech Embeddings using \\ Cross-modal Self-supervision}
\name{Arsha Nagrani$^1$*\thanks{\hspace{-12pt}* These authors contributed equally to this work.} , Joon Son Chung$^{1,2}$*\samethanks[1] , Samuel Albanie$^1$*\samethanks[1], Andrew Zisserman$^1$}
\address{$^1$ Visual Geometry Group, Department of Engineering Science, University of Oxford \\
$^2$ Naver Corporation \\ 
\small{\url{https://www.robots.ox.ac.uk/~vgg/research/cross-modal-disentanglement/}}
}
\begin{document}
%
\maketitle
\begin{abstract}
The objective of this paper is to learn representations of speaker identity without access to manually annotated data.  To do so, we develop a self-supervised learning objective that exploits the natural cross-modal synchrony between faces and audio in video.   The key idea behind our approach is to tease apart---without annotation---the representations of linguistic content and speaker identity.  We construct a two-stream architecture which: (1) shares low-level features common to both representations; and (2) provides a natural mechanism for explicitly disentangling these factors, offering the potential for greater generalisation to novel combinations of content and identity and ultimately producing speaker identity representations that are more robust.

We train our method on a large-scale audio-visual dataset of talking heads `in the wild', and demonstrate its efficacy by evaluating the learned speaker representations for standard speaker recognition performance. 
\end{abstract}


\section{Introduction}

The coupling of deep neural networks with large-scale labelled training datasets has produced a number of notable successes, yielding improved performance in speech related tasks such as ASR~\cite{chiu2018state} and speaker verification~\cite{torfi2018text,xie2019utterance}.  However, the considerable cost of manually producing such labels ultimately limits the potential of fully supervised approaches.  By contrast, methods which are able to learn effective representations from data with few labelled examples can in principle benefit from the ever-increasing quantity of existing unlabelled speech data.

%
 \begin{figure}[ht]
\begin{center}
\includegraphics[width=0.4\textwidth]{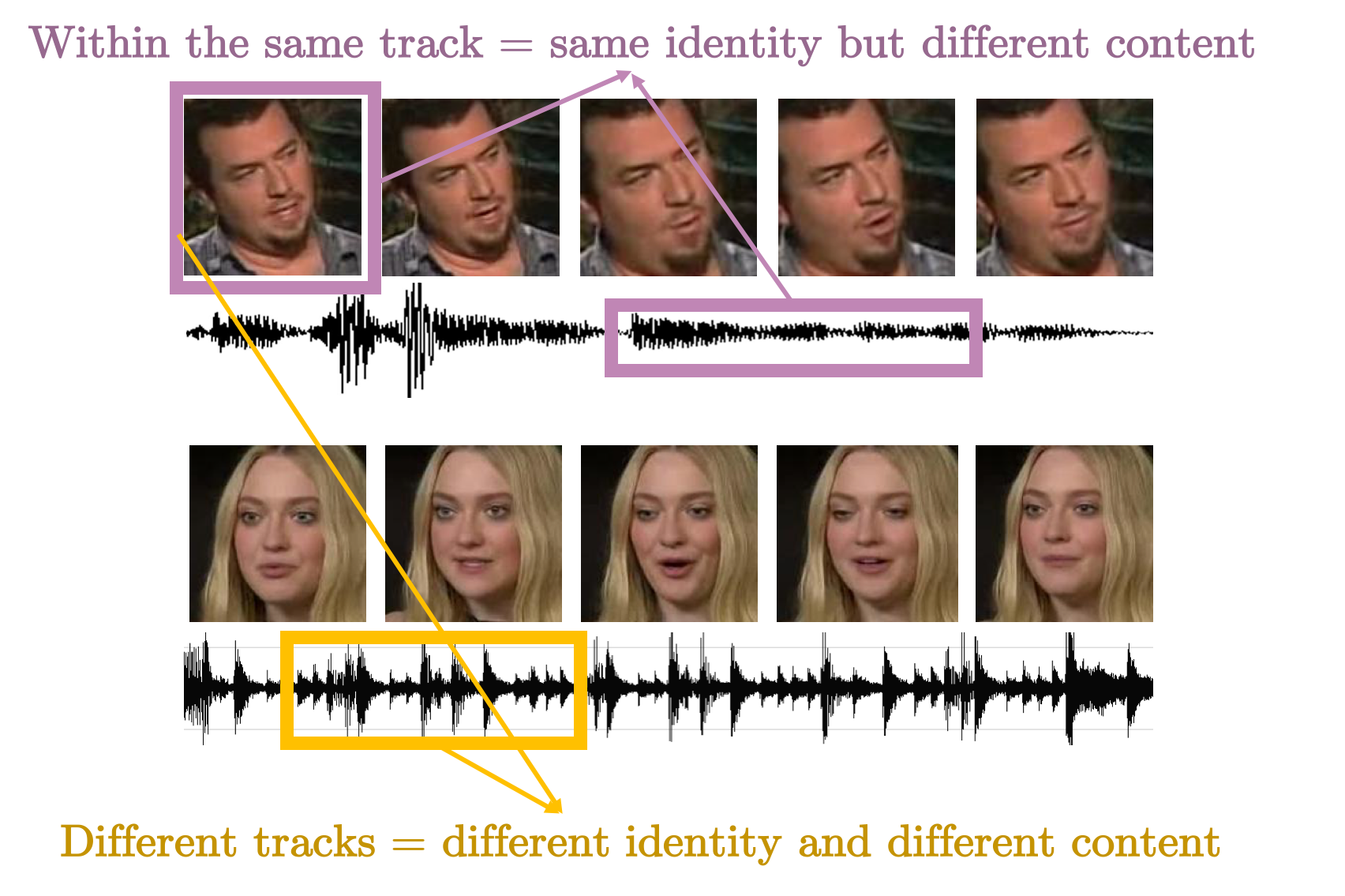}
\end{center}
\vspace{-20pt}
   \caption{To learn representations for speaker recognition without labels, our method relies on two hypotheses: (1) face and voice samples within a single face-track are likely to share a common identity, but different linguistic content across time; (2) face and voice samples from different face-tracks are likely to have both different speaker identities and different linguistic content.}
   
   
\label{fig:teaser}
\vspace{-5pt}
\end{figure}
The objective of this paper is to develop one such method for learning compact and robust representations of speaker identity without supervision.  Ultimately, these representations can then be used for a number of downstream tasks such as speaker recognition, clustering and diarisation etc.  To achieve this goal, we propose to exploit the natural synchrony between faces and audio in audio-visual video data as a supervisory signal, removing the need for speaker annotation. To facilitate our method, we assume access to a large-scale collection of unlabelled speaking face-tracks \cite{Nagrani17}, which can be readily obtained through self-supervised techniques for active speaker detection \cite{chung2016out}.  Beyond access to this data, our approach makes use of two weak statistical cues to define a self-supervised learning objective (Fig. 1): we assume that faces and voices extracted within a face-track at small offsets are likely to have the same speaker identity but different linguistic content, while faces and voices from different face-tracks are likely to differ in both content and speaker identity. As we show in Sec.~\ref{sec:model}, these cues can be combinedd to learn representations of speaker identity which minimise their dependence on speaker content. The motivation for doing so is simple: unlike earlier datasets such as TIMIT~\cite{garofolo1993darpa} that are carefully balanced for phonetic and dialectal coverage, more modern (and larger) datasets created from uncontrolled speech `in the wild' are likely to contain a strong correlation between identity and linguistic content. For example, VoxCeleb2~\cite{chung2018voxceleb2} consists of interviews of famous celebrities from a wide variety of professions, whose speech can be closely tied to their occupation---the cricketer Adam Gilchrist says the word \textit{`cricket'} 17 times and \textit{`president'} 0 times;  whereas the politician Nancy Pelosi says the word \textit{`president'} 88 times, \textit{`Democrats'} 19 times and \textit{'cricket'} 0 times. Consequently, a model trained to represent identity may be incentivised to use linguistic content as a discriminative cue.  While some coupling between content and identity is natural, over-reliance on content can prevent generalisation to new settings, harming robustness. More broadly, disentangled representations can, in principle, achieve an exponential improvement in generalisation efficiency over their entangled counterparts, because they are able to represent novel combinations of factors that were encountered separately (but never in combination) during training. 


In this work, we make the following contributions:
(1)~We propose a novel framework for learning speech representations capturing information at different time scales in the speech signal, including in particular the identity of the speaker; (2)~we show that we can learn these representations from a large, unlabelled collection of \say{talking faces} in videos as a source of free supervision, without the need for any manual annotation; (3)~we show that sharing a trunk architecture for two different tasks (content and speaker identity) and adding an explicit disentanglement objective between the two improves performance; and, (4)~we evaluate the performance of our self-supervised embeddings on the popular VoxCeleb1 speaker recognition benchmark and compare to fully supervised methods. All data and models will be released. 





\input{related.tex}
\vspace{-1em}
\input{model.tex}

\vspace{-1em}
\input{experiments.tex}

\vspace{-1em}
\input{discussion.tex}

\vspace{5pt}\noindent\textbf{Acknowledgements}
This work is funded by the EPSRC Programme
Grant Seebibyte EP/M013774/1 and ExTol EP/R03298X/1. Arsha is funded by a Google PhD Fellowship.
\clearpage
\bibliographystyle{IEEEbib}
\bibliography{shortstrings,mybib}

\end{document}

%% file: related.tex
\section{Related Work}
\noindent\textbf{Representation Learning.} The ability to represent variable-length high-dimensional audio segments using compact, fixed-length representations
has proven useful for many speech applications such as speaker verification~\cite{xie2019utterance, chung2018voxceleb2}, audio emotion classification~\cite{albanie2018emotion}, and spoken term detection (STD)~\cite{miller2007rapid}, where the representation can be coupled with a standard classifier. The use of fixed-length representations also enables efficient storage and retrieval when paired with an inverted index. These representations can either be hand-crafted, such as MFCCs or learned from data - such as i-vectors and deep neural networks. While the former may fail to capture the correct underlying factors for the task, the latter require large amounts of expensively labeled training data to be effective.  As a consequence, there has recently been renewed interest in learning unsupervised audio representations~\cite{chung2016audio}.

\vspace{5pt}\noindent\textbf{Disentangled Representation Learning.} Motivated by their attractive compositional properties and theoretical ability to generalise efficiently, a number of models that seek to learn disentangled representations in a weakly supervised or self-supervised manner have been proposed, such as
 DC-IGN ~\cite{kulkarni2015deep}, InfoGAN~\cite{chen2016infogan} and VQ-VAE~\cite{van2017neural}. 
Due to the proliferation of video data, there has also been a renewed interest in learning representations from sequential data~\cite{fabius2014variational,chung2015recurrent,chung2016hierarchical,fraccaro2016sequential}. These self-supervised works focus on predicting future, missing or contextual information, all within the same modality. However to the best of our knowledge, no prior method has sought to learn disentangled representations through cross-modal self-supervision. 


\vspace{5pt}\noindent\textbf{Audio-Visual Self Supervision.} A number of recent works~\cite{aytar2016soundnet,arandjelovic2017look,aytar2017see,owens2016ambient, Korbar18} have explored the concept of exploiting the correspondence between
synchronous audio and visual data in teacher-student style
architectures (where the `teacher' is represented by a pretrained network)~\cite{aytar2016soundnet,aytar2017see}, or two-stream networks where both networks are trained from scratch~\cite{arandjelovic2017look,chung2016out}. Additional work has examined cross-modal
relationships between faces and voices specifically in order to learn identity~\cite{Nagrani18a,nagrani2018learnable,kim2018learning} or emotion~\cite{albanie2018emotion} representations. In contrast to these works, we aim to learn representations of both content and identity with a view to explicitly disentangling separate factors---we compare our approach with theirs in Sec.~\ref{sec:experiments}.

%% file: model.tex
\begin{figure}[ht]
\begin{center}
\includegraphics[width=0.4\textwidth]{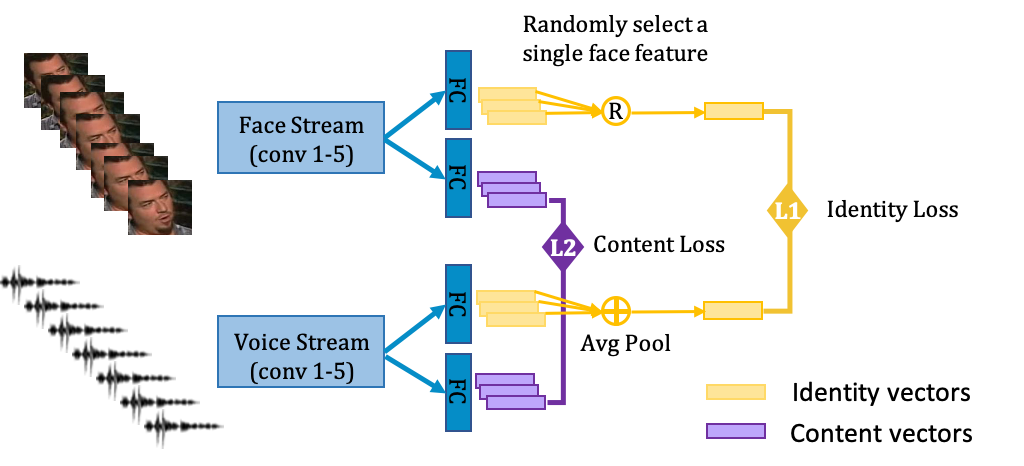}\\
\includegraphics[width=0.4\textwidth]{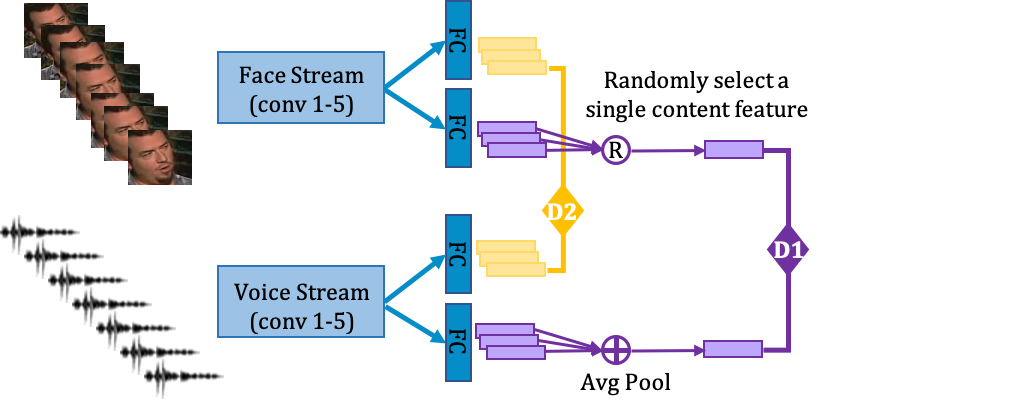}
\end{center}
\vspace{-20pt}
  \caption{The framework for learning representations of speaker identity. We aim to explicitly disentangle speech representations into content and identity embeddings. Under the notation of Sec.~\ref{sec:model}, we show $B=1$ facetracks in the input and $N=3$ samples per track. Top diagram:  the identity (L1) and content classification losses (L2); bottom diagram: the disentanglement losses (D1 and D2).}
  \vspace{-5pt}
\label{fig:model}
\end{figure}

\section{Model} \label{sec:model}

Speech, like many sequential natural signals, can be decomposed into the interaction of several largely-independent causal factors which express themselves over different time scales.  The central observation that underpins our approach is that the speaker identity affects fundamental frequency, pitch and volume at the utterance level while linguistic content affects spectral contour and duration of formants more locally. 

Without labels, we have no way to directly separate these factors.  Instead, we can impose our prior knowledge as to how such representations should behave.  Intuitively, representations of identity should change \textit{slowly} over time (remaining constant for a given speaker), whereas representations of content should change \textit{quickly}, capturing the local variation in the speech signal.  Concretely, we enforce these properties by exploiting the known correspondence between a speech signal and the face of its speaker within a facetrack to impose three constraints on the representations for content and identity: 

\vspace{5pt}\noindent\textbf{Content constraints.} Within a given speaking facetrack, speech and face signals extracted concurrently contain redundant (or overlapping) linguistic content (while this information is trivially available in the speech signal, it is perhaps less obvious that it is also present in the face---in fact, it is this signal that enables lipreading). By contrast, the face signal at a small temporal offset from the speech signal is likely to convey \textit{different} linguistic content.  These cues provide a natural source of paired data (positive and negative examples) that we can use to learn a self-supervised representation of linguistic content from a speech signal~\cite{chung2016out}. 

\vspace{5pt}\noindent\textbf{Identity constraints.} By considering instead face and voice signals across face tracks, we can obtain a different form of constraint: signals from the same face-track should come from the same speaker, while those from different face-tracks are likely to come from different speakers. This idea was demonstrated in ~\cite{nagrani2018learnable}. 

\vspace{5pt}\noindent\textbf{Disentangling constraint.} Although representations that have been trained to satisfy the intra-track and inter-track constraints may capture a measure of both linguistic content and speaker identity, there is no guarantee that both factors will be disentangled (represented independently of one-another). To achieve this last goal, we employ a further constraint on the speech representations themselves, requiring that variation within one factor cannot be predicted from the other to enforce their independence. \\

\vspace{3pt}\noindent\textbf{Learning framework.}
In this work, we train a single model in an end-to-end self-supervised manner to satisfy the constraints described above (the framework is depicted in Fig.~\ref{fig:model}). In the next section, we describe the architecture used for representation learning and the losses that are used to implement these constraints. All losses are across modalities.


\subsection{Network Architecture}
\vspace{-1mm}
Our architecture consists of two sub-networks, one sub-network that ingests five cropped faces as input,  and another sub-network that takes in short-term magnitude spectrograms of 0.2-second speech segments. Each sub-network contains a block of five convolutional layers as the basic feature extraction trunk (these are shared for both content and identity, as it has been speculated that lower level features, e.g.\ edges for images and formants for speech, are likely to be common~\cite{shinohara2016adversarial} for different high level tasks). Both sub-networks are based on the VGG-M architecture \cite{Chatfield14} which strikes a good trade-off between efficiency and performance. See \cite{chung2019perfect} for the exact filter sizes. 
After this, each block branches into two separate fully connected layers, one that produces identity embeddings and one that produces content embeddings, both of dimension 1024. For $N+4$ input frames, $N$ identity and content embeddings are produced for each modality stream (Fig.~\ref{fig:model}), since both sub-networks have temporal receptive fields of 5 frames (0.2 second) and strides of 1 frame (0.04 second).
During training, the identity vectors from the audio stream are then averaged into a single vector, while a single identity vector is selected from the face stream at random. To understand this choice, note that if we were to also average the face embeddings, then the task of matching identity representations would simply become one of lip reading, i.e.\ matching the linguistic content of the audio and visual signals. Hence we pick a single random face vector and make the assumption that a face from a single frame is insufficient to encode linguistic content. 

\vspace{5pt}\noindent\textbf{Self-Supervised Paired Data Inputs.}
In a single minibatch, we take $B$ face-tracks, each of 1.2 seconds. Within a face-track, we sample $N+4$ consecutive face images and $N+4$ temporally aligned speech segments from the 1.2-second speech segment. Hence the total number of input samples per batch is $(N+4) \times B$ face images and $(N+4) \times B$ speech segments. 
\subsection{Loss Functions}
\vspace{-1mm}
\noindent A \textit{content loss (CL)} is used to implement the content constraint via a multi-way matching task, as described  in~\cite{chung2019perfect}.  The loss takes one input feature from the visual stream and $N$ features from the audio stream. Since only one of these audio features is a positive sample (i.e. in sync with the visual stream), this can be set up as any ($N$)-way feature matching task. Euclidean distances between the audio and video features are computed, resulting in $N$ distances. A cross-entropy loss is applied on the inverse of this distance after passing through a softmax, encouraging the similarity between matching pairs to exceed that of non-matching pairs. 

\vspace{5pt}\noindent An \textit{identity loss (IL)} is used to implement the identity constraint.  It is similar in form to the content loss, but the negative samples are now obtained from different tracks, as opposed to \textit{within} a track. The task becomes one of selecting the correct track averaged identity speech representation for a single face representation from all the $B$ tracks in a batch, i.e. this is a B-way classification task. 

\vspace{5pt}\noindent \textit{Disentanglement losses (DL)} are used to encourage explicit separation of representations---for this we use the confusion loss implemented by~\cite{alvi2018turning} (inspired by~\cite{tzeng2015simultaneous}). This loss is used to assess the amount of spurious variation information left in either feature representation and then remove it (for the identity representation, content information is a spurious variation and vice versa). Minimizing this loss seeks to change the feature representation, such that it becomes invariant to the spurious variations. To remove identity from content, we perform the B-way identity matching task \textit{across} facetracks using the content vectors as input instead (D1 in Fig.~\ref{fig:model}). 
We then minimise the cross-entropy between the output predicted from the model and a uniform distribution. Similarly, we apply the N-way content classification loss to the identity vectors and minimise the cross-entropy with the output to a uniform distribution (D2 in Fig.~\ref{fig:model}). See~\cite{alvi2018turning}, Equations 1--3 for exact details.


%% file: experiments.tex
\section{Experiments} \label{sec:experiments}
We train our model using the following loss combinations: (1) Only the content loss: in this case the identity streams are not present in the network; (2) Using only the identity loss: in this case the content streams are not present in the network; (3) Joint training with both the content and the identity loss; (4) Joint training with the content, identity and disentanglement losses.  In all cases the model uses the same trunk architecture and training hyperparameters.

\vspace{5pt}\noindent\textbf{Implementation Details.}
The model is implemented using PyTorch. It is trained end-to-end with batch size $B=30$ and $N=30$ samples per face-track using SGD (initial learning rate of \texttt{1e-2} which decays by $0.95$ per epoch). 

\subsection{Dataset} \label{sec:data}
We train our model on VoxCeleb2~\cite{chung2018voxceleb2}, a large-scale audio-visual dataset of interviews obtained from unedited YouTube videos. The dataset consists of over a million utterances for 6,112 identities.  No identity labels are used during training. To reduce computational cost, we sample only 20\% of the speech per speaker for training from the VoxCeleb \textit{dev} set, and validate performance of the self-supervised learning objectives on $120$ speakers from the VoxCeleb2 test set. The statistics of the dataset can be seen in Table \ref{table:data}. 


\begin{table}[ht]

\centering 
\begin{tabular}{l c c } 
\toprule
 & \textbf{\# face-tracks} & \textbf{\# identities} \\ 
\midrule 
Training set & 218,340 & 5,994 \\
Test set & 36,600 & 120 \\
\bottomrule
\end{tabular}
\caption{Dataset Statistics. Although we report the no.\ of identities in the dataset, the identities are \textit{not used at any point during training}.}
\label{table:data} 
\end{table}
\subsection{Evaluation} 
We first evaluate the performance of our model on the two self-supervised learning objectives that it was trained for, and then evaluate the learned representations on the downstream task of speaker recognition on the standard VoxCeleb1 speaker recognition benchmark.

\begin{table}[ht]
\centering 
\setlength{\tabcolsep}{4pt}
\begin{tabular}{l c c c } 
\toprule
& \textbf{Content Task} & \multicolumn{2}{c}{\textbf{Identity Task}}  \\ 
\midrule
& $N$-way cls. & $B$-way cls. & EER \\ 
\midrule
Random & 3.3\% & 3.3\% & 50.0\%\\ 
Content loss only & 49.0\% & -- & -- \\
Identity loss only & --
& 44.3\% & 24.8\% \\ 
\midrule
\multicolumn{4}{l}{\textit{Content Embeddings}}\\
Con. and Id. Loss & 46.7\% & 8.5\% & 45.7\% \\
Con., Id. and Dis. Loss & 49.0\% & 10.5\% & 45.2\% \\
\midrule
\multicolumn{4}{l}{\textit{Identity Embeddings}}\\
Con. and Id. Loss  & 19.3\% & 48.2\% & 23.1\% \\ 
Con., Id. and Dis. Loss & \textbf{12.0\%} & \textbf{49.6\%} & \textbf{18.9\%} \\ 
\bottomrule
\end{tabular}
\caption{Results on the self-supervised training objectives. The content task is $N$-way classification ($N$ = number of samples per face-track), and the Identity task is $B$-way classification ($B$ = number of face-tracks per minibatch). With $N=B=30$, random performance is 3.3\%. Lower EER, higher cls.\ accuracy is better. We want good performance of identity embeddings on the identity task, and low performance on the content task.}
\label{table:objective} 
\vspace{-12pt}
\end{table}

\vspace{5pt}\noindent\textbf{Learning Objective.}  We evaluate the self-supervised learning objectives on $120$ speakers from the VoxCeleb2 test set (Table \ref{table:data}), and the results can be seen in Table \ref{table:objective}. We evaluate the learned identity representations on the N-way classification task within a facetrack (content task), as well as evaluating it on the identity B-way classification task. From Table \ref{table:objective}, it is clear that training both self-supervised objectives jointly improves performance on the identity classification task over training for identity alone (48.2 \% vs 44.3 \%) and training with the disentanglement losses provides a further improvement (49.6 \%). In order to further probe the effect of the disentanglement losses, however, we look at the performance of the identity embeddings on the content classification task (which ideally it should perform poorly on). From Table \ref{table:objective}, it can be seen that disentanglement helps remove content information from the identity embedding -- the accuracy drops from 19.3 \% to 12.0 \%,  on the N-way content classification task. 

As an aside, we also report performance of the content embeddings in the middle two rows of Table \ref{table:objective} (although learning content representations for their own sake is not the objective of this work) and note that joint training actually harms the performance compared to training with the content loss alone (from 49.0\% to 46.7\%) on the content classification task, however this performance is recovered by adding in the disentanglement losses. This is to be expected, as it is very difficult for identity information to leak into the content representation when it is trained for content alone (the content objective is trained with a large number of \textit{negative} pairs within the same face-track, discouraging the embedding from learning identity). 

\vspace{5pt}\noindent\textbf{Speaker Recognition.}
We then extract identity embeddings for the data in the VoxCeleb1 \textit{test} set (VoxCeleb1-O, 40 speakers)~\cite{Nagrani17}. We first evaluate using the self-supervised embeddings directly (i.e.\ without \textit{any} speaker identity labels at all), and report results in Table~\ref{table:results}. The negative cosine distance between embeddings is calculated directly and used as the similarity score between verification pairs. Once again we see a similar trend in the results, both joint training and disentanglement show cumulative gains in performance. 
We then compare our method to fully supervised performance, by freezing the layers of our network and then finetuning a single fully connected layer on the embedding network with n-pair loss, using labels from the VoxCeleb1 \textit{dev} set. We do this for various subsets of the VoxCeleb1 \textit{dev} set, and demonstrate in  Table~\ref{table:supresults} that up until 500 speakers, our self-supervised method (even with only the identity loss, and with gains using the other two losses) outperforms full supervision. The fully supervised baseline is trained end-to-end, and for a fair comparison, has the exact same architecture as the audio stream of the cross-modal model.

\begin{table}[ht]
\centering 

\begin{tabular}{l c }
\toprule
\textbf{Method} & \textbf{EER}  \\ 
\midrule
Identity loss only & 23.15\% \\
Identity loss + Content loss & 22.59\% \\
Identity loss + Content loss + Dis. loss & \textbf{22.09\%} \\
\bottomrule
\end{tabular}
\caption{Speaker verification results on the VoxCeleb1 test set. Lower is better. EER: Equal Error Rate.}
\label{table:results} 
\vspace{-12pt}
\end{table}

\begin{table}[ht]
\centering 
\setlength{\tabcolsep}{4pt}
\begin{tabular}{l c c c c}
\toprule
\textbf{\# speakers}  & 100 & 250 & 500 & 1,211 \\ 
\textbf{\# utterances} & 1,228 & 6,019 & 12,146 & ALL \\ 
\hline 
Id. loss only & 15.05\% & 13.00\% & 11.16\% & 9.85\% \\
Id.+Cont.+Dis. loss & \textbf{14.33\%} & \textbf{12.69\%} & \textbf{10.94\%} & 9.43\% \\
Fully supervised & 19.84\% & 13.60\% & 11.35\% & {\bf 7.28\%} \\
\bottomrule
\end{tabular}
\caption{Comparison to fully supervised performance on the VoxCeleb1 test set measured in EER. For the first two rows, a single fully connected layer is trained on the self-supervised embeddings. The fully supervised model is trained end-to-end with labels. Lower is better.}
\label{table:supresults} 
\vspace{-12pt}
\end{table}

%% file: discussion.tex
\section{Conclusion}

In this work we develop a self-supervised method that learns speaker recognition embeddings from speech without access to any training labels, simply by using the co-occurence of faces in video. By explicitly disentangling factors of variation such as content and identity, and training for both objectives with a common trunk architecture, we show improvements in generalisation to unseen speakers, and in the case of small amounts of training data, even outperform fully supervised methods. 